\documentclass[conference]{IEEEtran}
\usepackage{cite}
\usepackage[squaren, Gray, cdot]{SIunits}
\usepackage[usenames,dvipsnames,svgnames,table]{xcolor}

\ifCLASSINFOpdf
   \usepackage[pdftex]{graphicx}
\else
\fi
\usepackage{amsmath}
\DeclareMathOperator*{\POPCOUNT}{POPCOUNT}

\usepackage{array}

\begin{document}
\title{ Embracing the Unreliability of Memory Devices \\
for Neuromorphic Computing
\thanks{\textbf{This work was supported by the ERC Grant NANOINFER (715872) and  the ANR grant NEURONIC (ANR-18-CE24-0009).}}
}

\author{\IEEEauthorblockN{
Marc Bocquet\IEEEauthorrefmark{1}, 
Tifenn Hirtzlin\IEEEauthorrefmark{2}, 
Jacques-Olivier Klein\IEEEauthorrefmark{2}, 
Etienne Nowak\IEEEauthorrefmark{3}, \\
Elisa Vianello\IEEEauthorrefmark{3}, 
Jean-Michel Portal\IEEEauthorrefmark{1} 
and Damien Querlioz\IEEEauthorrefmark{2}}
\IEEEauthorblockA{\IEEEauthorrefmark{1}IM2NP, Univ. Aix-Marseille et Toulon, CNRS, France.}
\IEEEauthorblockA{\IEEEauthorrefmark{2}Universit\'e Paris-Saclay, CNRS, C2N, 91120 Palaiseau, France.  Email: damien.querlioz@c2n.upsaclay.fr}
\IEEEauthorblockA{
\IEEEauthorrefmark{3}Universit\'e Grenoble-Alpes, CEA, LETI, Grenoble, France.}
\IEEEauthorblockA{\textit{Invited Paper}}}
\IEEEoverridecommandlockouts
\maketitle
\begin{abstract}
The emergence of resistive non-volatile memories opens the way to highly energy-efficient computation near- or in-memory. However, this type of computation is not compatible with conventional ECC, and has to deal with device unreliability. Inspired by the architecture of animal brains, we present a manufactured differential hybrid CMOS/RRAM memory architecture suitable for neural network implementation that functions without formal ECC. We also show that using low-energy but error-prone programming conditions only slightly reduces network accuracy.
\end{abstract}


%
\IEEEpeerreviewmaketitle

\section{Introduction}

Emerging nonvolatile memory technologies such as resistive, phase change and spin torque magnetoresistive memories offer considerable opportunities to advance microelectronics, as these memories are  faster than flash memories, while being compact and compatible with the integration in the backend-of-line of modern CMOS processes \cite{ielmini2018memory,yu2018neuro}. 
However, although these technologies are usually more reliable than flash memories, they remain considerably less reliable than volatile charge-based random access memories. 
Strategies for reducing errors due to device variation and limited endurance involve costly materials and technology developments \cite{bayat2018implementation}, 
energy-consuming special programming strategies \cite{chang201419}, 
and quite universally, the reliance on advanced multiple error correcting codes (ECC) \cite{ielmini2018memory,golonzka2018mram}, requiring large area and energy hungry decoding circuitry \cite{gregori2003chip}. 

The existence of errors in emerging memories is also a severe limitation for the development of in or near-memory computing schemes, which aim at achieving highly energy efficient computation by eliminating the von Neumann bottleneck \cite{editorial_big_2018,ielmini2018memory}. 
In or near-memory computing schemes are indeed hardly compatible with ECC, as computation is performed with multiple row selection or in the sensing circuit \cite{bocquet2018,hirtzlin2019digital}. These constrains are in sharp contrast with animal brains, which function with vastly unreliable, redundant, memory devices (synapses) without using formal error correction \cite{faisal2008noise,klemm2005topology}. 

In this work, we show through an example that in computing architectures inspired by brains (neuromorphic architectures), memory device variability can  to a large extent be ignored, and even embraced, and that this attitude can provide important benefits. We first present a differential memory architecture optimized for the ECC-less in-memory implementation of biarized neural networks. We show based on experimental measurements on a fabricated CMOS/RRAM hybrid chip and on network simulations that this architecture can mostly ignore device variation, and investigate the benefits of accepting errors. Based on a modeling study, we show that the same methodology could be transferred to MRAM.


\section{An In-Memory Computing Memory Block That Functions With Error-Prone Devices}
\label{sec:background}

\begin{figure}[ht]
	\centering
	\vspace{0.5cm} 
	\includegraphics[width=3.45in]{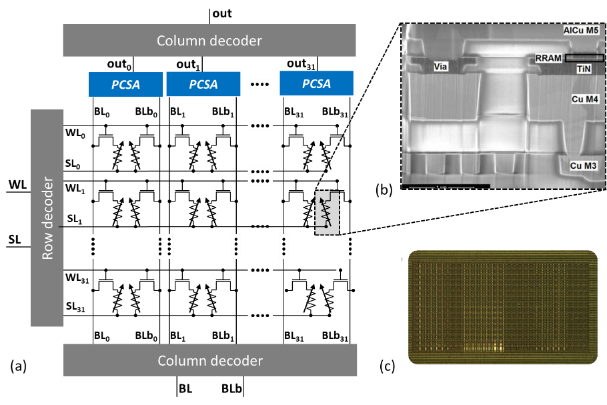}
	\caption{(a) Simplified schematic of our in-memory computing hybrid CMOS/RRAM test chip. 
	(b) Electron microscopy image of an RRAM cell integrated in the  backend-of-line of a 130~nm commercial CMOS technology. 
	(c) Photography of the die.}
	\vspace{0.5cm} 
	\label{fig:archi}
\end{figure}

\begin{figure}[ht]
    \vspace{0.5cm} 
	\centering
	\includegraphics[width=3.45in]{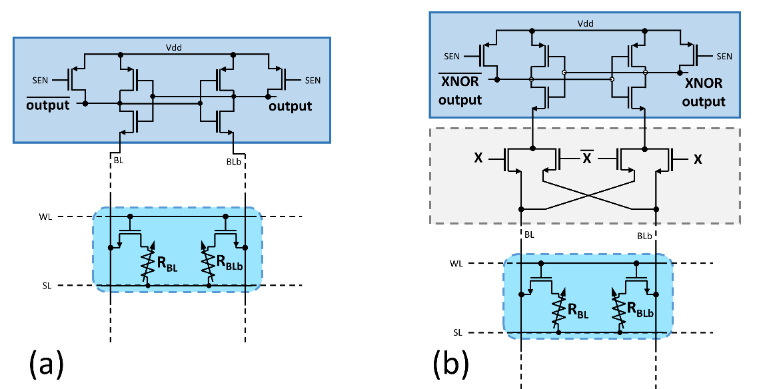}
	\caption{Circuit of the precharge sense amplifier (PCSA) used in the test chip of Fig.~\ref{fig:archi}. 
	(a) Standard version, (b) version augmented with XNOR operation, initially proposed in \cite{zhao2014synchronous}. }
	\vspace{0.5cm} 
	\label{fig:PCSA}
\end{figure}

\begin{figure}[ht]
	\centering
	\vspace{0.5cm}
	\includegraphics[width=3.45in]{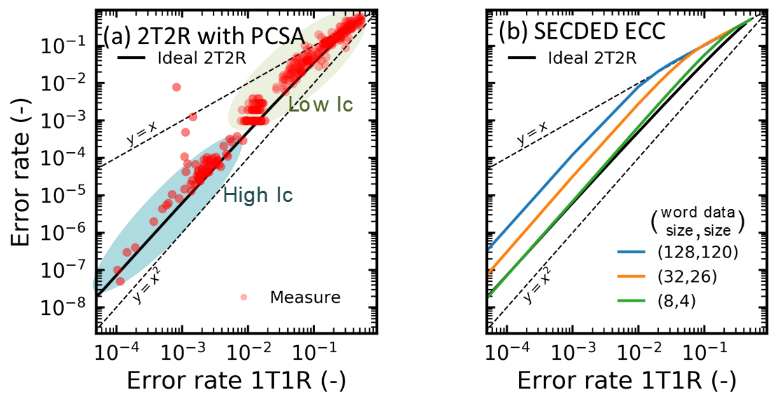}
	\caption{(a) Bit error rates (BER) measured using the PCSAs of the test chip as a function of 1T1R BER in the same conditions. 
	(b) For comparison, improvements of BER obtained using standard Single Error Correction Double Error Detection (SECDED) ECC. Figure adapted from \cite{hirtzlin2019digital}.}
	\vspace{0.5cm}
	\label{fig:ECC}
\end{figure}

In this work, we propose the use of a memory architecture where each bit is stored in a two-transistor/two-resistor (2T2R) cell. 
We implemented a kilobit version (2,048 devices) of this architecture in a 130~nm CMOS technology, with hafnium oxide-based RRAM fully embedded in the backend-of-line (Fig.~1). This test chip was initially introduced in \cite{bocquet2018,hirtzlin2019digital}. 
Bits are stored in a differential fashion between the two devices to reduce errors. 
Doing so, during the read phase, a high resistive state (HRS) is always compared to a low resistive state (LRS), doubling the memory read window with regards to the conventional comparison to a reference value between HRS and LRS, as is used in  one-transistor/one-resistor (1T1R) architectures \cite{bocquet2018}. 
This differential read scheme is operated by on-chip precharge sense amplifiers (PCSA), whose circuit is presented in Fig. 2(a). 
These sense amplifiers can also be augmented to directly perform logic operations during read operations \cite{zhao2014synchronous}. An example where a PCSA has been augmented to perform exclusive NOR (XNOR) operation is shown in Fig.~2(b). Such in-memory computing augmentations, while approaching logic and memory, make our system incompatible with conventional ECC scheme.

Extensive experimental measurements on our test chip showed that the 2T2R strategy indeed reduces bit errors  when compared to the classical 1T1R approach.  
RRAM devices error rate is directly linked to the current used during the programming operations, offering a knob of error rate tuning depending on the application requirements. 
Fig.~3(a) compiles statistical measurements on the fabricated test chip, taken with diverse programming currents, allowing evaluating the bit error rates (BER) benefits of the 2T2R approach in different conditions. It is apparent in this Figure that the 2T2R strategy always reduces the amount of bit errors, with the highest benefits seen at lower BERs. The detailed methodology for obtaining  Fig.~3(a)  is presented in \cite{hirtzlin2019digital}.

Quite interestingly the error reduction benefits of the 2T2R approach are similar to the one of a Single Error Correcting Double Error Detecting  ECC (SECDED, or extended Hamming), but without the high peripheral circuit overhead required by this ECC \cite{gregori2003chip}, and associated read performance degradation (Fig. 3(b)). Moreover, this result is obtained considering the same memory capacity (2T2R without ECC versus 1T1R plus extra bit for correction code storage).


\section{Benefits at the Network Level}
\label{sec:circuit}

\begin{figure}[ht]
	\centering
	\vspace{0.5cm}
	\includegraphics[width=3.45in]{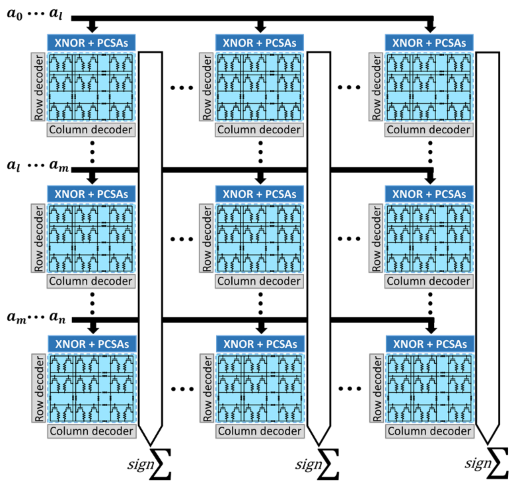}
	\caption{Schematization of a full digital system implementing a Binarized Neural Networks using in-memory computing blocks of Fig.~\ref{fig:archi}.}
	\vspace{0.5cm}
	\label{fig:fullarchi}
\end{figure}

\begin{figure}[ht]
	\centering
	\vspace{0.5cm}
	\includegraphics[width=3.45in]{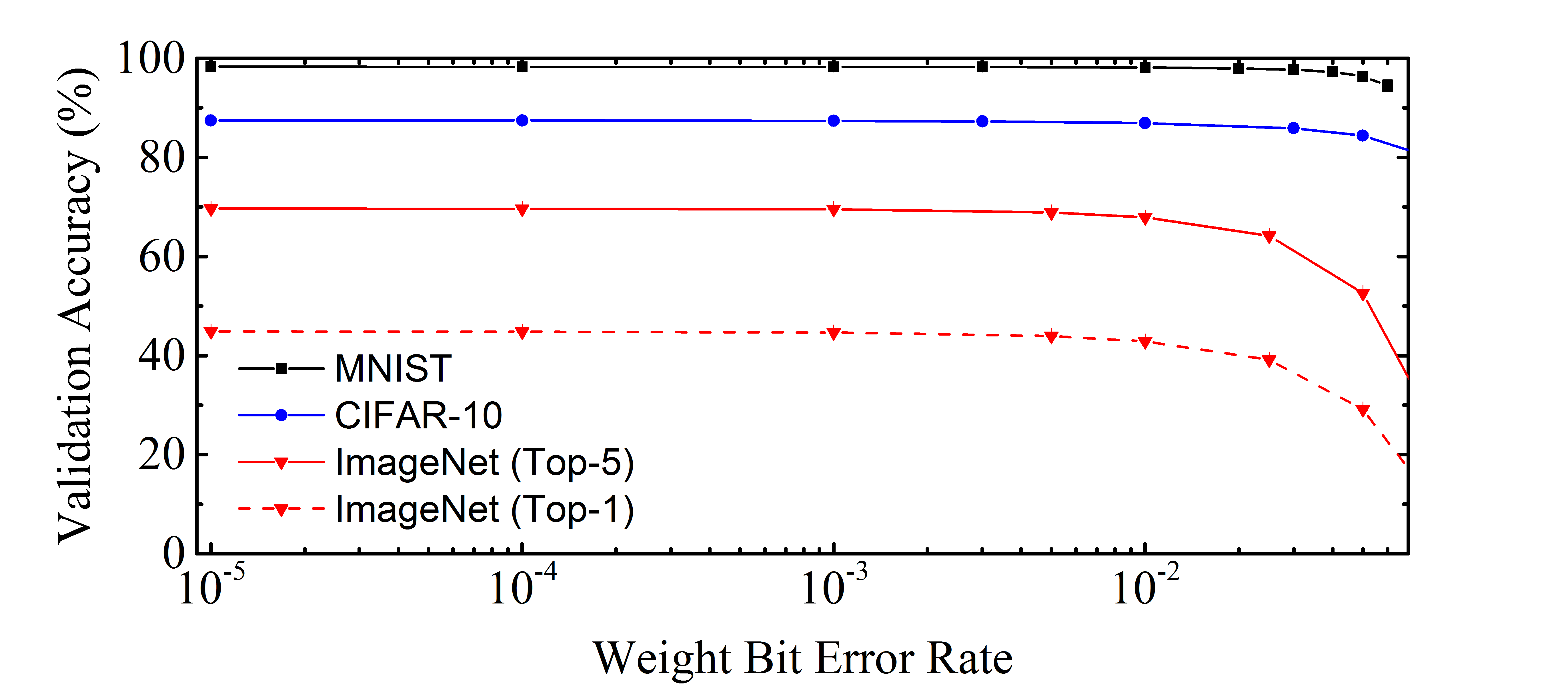}
	\caption{ Impact of the BER of memories on applications of Binarized Neural Network: handwritten digit recognition (MNIST), image recognition (CIFAR-10, ImageNet TOP-1 and TOP-5). Details about the neural network architectures are provided in \cite{hirtzlin2019digital}.}
	\vspace{0.5cm}
	\label{fig:BER_impact}
\end{figure}

\begin{figure}[ht]
	\centering
	\vspace{0.5cm}
	\includegraphics[width=3.45in]{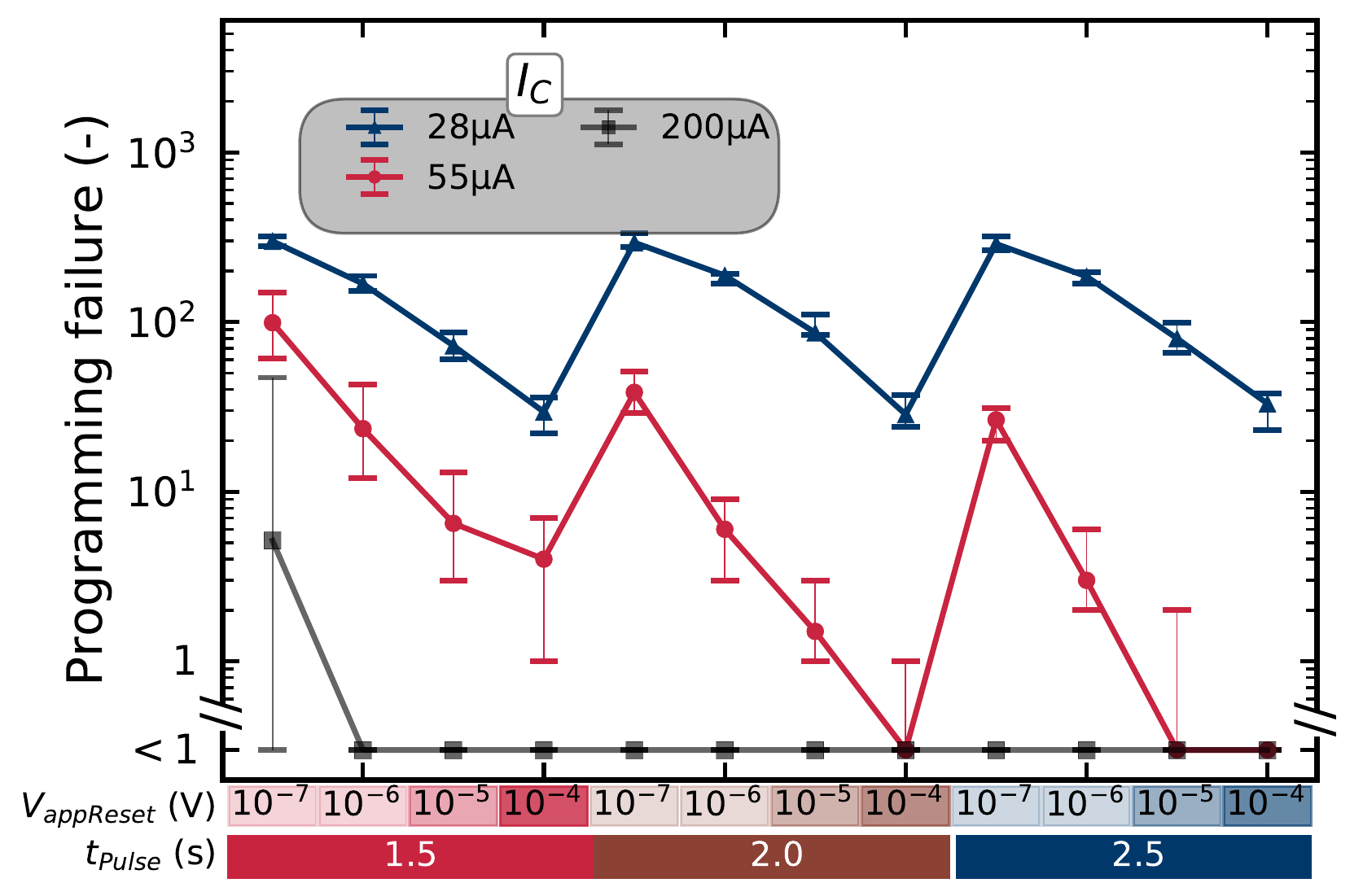}
	\caption{Number of errors on a one kilobit array using the 2T2R strategy (with PCSA) for different programming conditions (compliance current $I_C$, RESET voltage $V_{appReset}$, and programmming pulse duration $t_{pulse}$). Error bars represent the minimum and the maximum number of errors over five trials of the experiment. Figure adapted from \cite{hirtzlin2019digital}.
	}
	\vspace{0.5cm}
	\label{fig:programmingconditions}
\end{figure}

\begin{figure}[ht]
	\centering
	\vspace{0.5cm}
	\includegraphics[width=3.45in]{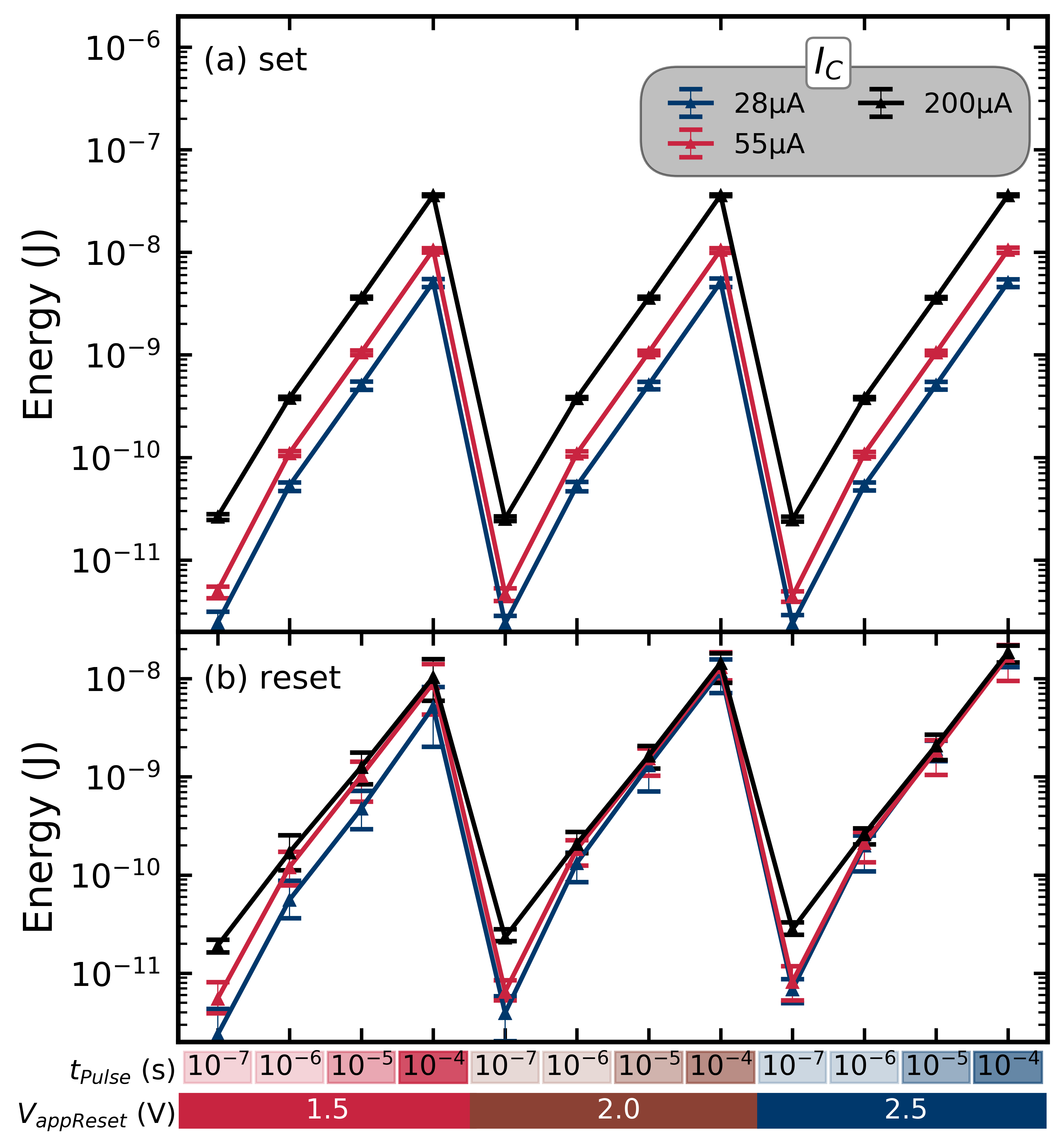}
	\caption{Mean programming energy (per bit) of RRAM cells in the (a) SET and (b) RESET processes for the programming conditions shown in Fig.~6.
	}
	\vspace{0.5cm}
	\label{fig:programmingconditions_energie}
\end{figure}

\begin{figure}[ht]
	\centering
	\vspace{0.5cm}
	\includegraphics[width=3.45in]{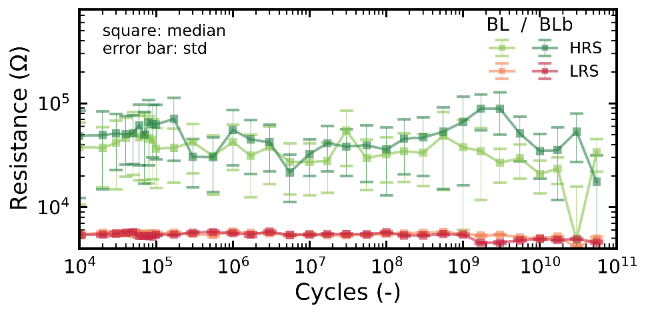}
	\caption{Endurance measurement for two devices (bit line BL and bit line bar BLb), programmed in weak conditions ($V_{appReset}=1.5V$, $I_C=200\mu A$, $t_{pulse}=1\mu s$). Figure adapted from \cite{hirtzlin2019digital}.}
	\vspace{0.5cm}
	\label{fig:cycling}
\end{figure}

\begin{figure}[ht]
	\centering
	\vspace{0.5cm}
	\includegraphics[width=3.45in]{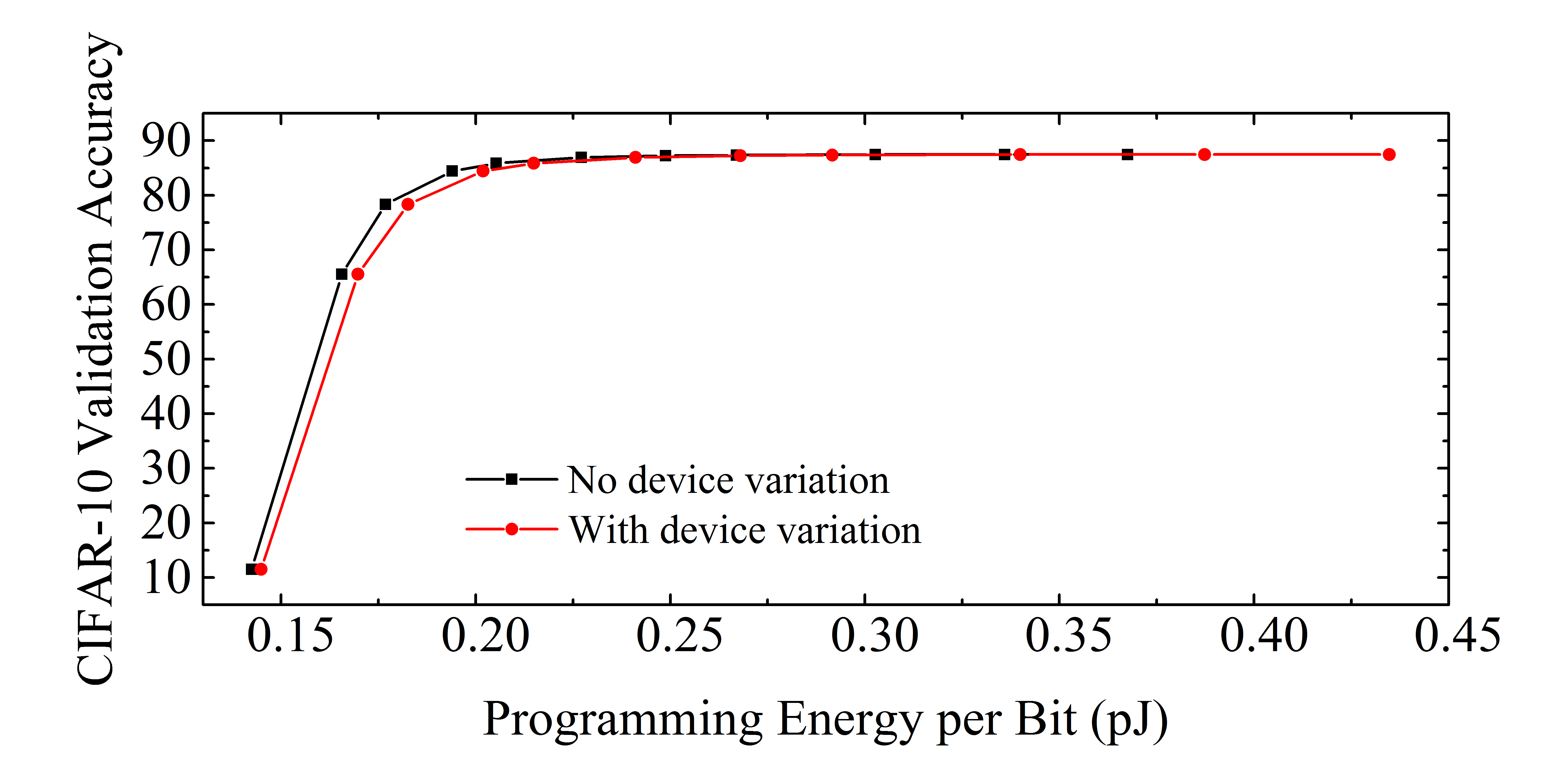}
	\caption{Accuracy on the CIFAR-10 image recognition task of a 28nm-technology MRAM based Binarized Neural Network, as a function of MRAM programming energy (varying programming time). Computed using the model of \cite{hirtzlin2019implementing}, considering or ignoring MOSFET and magnetic tunnel junction device variation.}
	\vspace{0.5cm}
	\label{fig:MRAM}
\end{figure}

Binarized Neural Networks (BNNs) \cite{courbariaux2016binarized}, or the highly similar XNOR-NETs  \cite{rastegari2016xnor},
are a recently proposed type of neural network, where synaptic weights and neuron states can take only binary values (meaning $1$ and $-1$) during inference (whereas these parameters assume real values in conventional neural networks).
Therefore, the equation for the activation $A$ of a neuron in a conventional neural network
\begin{equation}
\label{eq:activ_real}
    A =  f \left( \sum_i W_iX_i \right),
\end{equation}
(where $X_i$ are  inputs of the neuron,  $W_i$ the synaptic weights and $f$ its nonlinear activation function) simplifies into
\begin{equation}
\label{eq:activ_BNN}
    A = \mathrm{sign} \left( \POPCOUNT_i \left( XNOR \left( W_i,X_i \right) \right)-T \right).
\end{equation}
$\POPCOUNT$ is an integer function that counts the number ones. $\mathrm{sign}$ is the sign function, and $T$ is the threshold of the neuron, obtained during training by the use of the batch-normalization technique \cite{ioffe2015batch}.  

BNNs can achieve surprisingly high accuracy in vision  \cite{rastegari2016xnor,lin2017towards} or signal-processing \cite{bogdan} tasks.
BNNs have highly reduced memory requirements with regards to real neural networks, and have the added benefit of not requiring any multiplication, as this operation  is replaced by XNOR logic operations.
These advantages make BNNs  outstanding candidates for in-memory computing \cite{yu2016binary,giacomin2019robust,sun2018fully,sun2018xnor,tang2017binary,zhou2018new,bankman2018always,chang2019nv}.

The architecture of Fig.~1 is particularly adapted for the ECC-less implementation of such neural networks. 
For example, Fig.~4 shows a full system using memory circuits of Fig.~1 to implement a BNN.
The architecture uses the sense amplifier of Fig.~2(b) \cite{zhao2014synchronous} to implement XNOR operations directly in each memory circuit during the read phase, whereas the $\POPCOUNT$ operation, as well as neuron activation  are performed on foot of array columns using fully digital circuits. 
Refs.~\cite{hirtzlin2019digital,hirtzlin2019stochastic} describe this architecture in detail, as well as some its variations, and show that this architecture features outstanding energy-efficiency properties. 

We now evaluate the impact of errors in memories in this architecture. Fig.~5 shows simulations of the architecture programmed to perform several tasks:
the classic MNIST handwritten digit recognition task \cite{lecun1998gradient},
the CIFAR-10 image classification task \cite{krizhevsky2009learning},
and the  challenging  ImageNet classification task, which consists in classifying high-resolution images into 1,000 classes \cite{krizhevsky2012imagenet}.
The detailed architecture  of the BNNs used on these three tasks in presented in \cite{hirtzlin2019digital}.
All these tasks were simulated with various bit error rates on the memory devices.
Quite astonishingly, we see on all these three tasks
that bit error rate as high as $10^{-3}$ can be tolerated with little consequence on the accuracy of the implemented neural network.  This highlights that when implementing BNNs, memory perfection is far from being required.
Some dedicated training strategies could enhance this error tolerance even further \cite{hirtzlin2019outstanding}.

The combination of the fact that the 2T2R approach allows reducing the amount of bit errors, and that the BNN application features inherent tolerance to bit errors has important consequences in practice. 
It allows us to use RRAM devices in regimes where they are extremely unreliable. This can provide important energy savings: we can use devices with very weak programming conditions (low current and voltages, short programming time), where they feature high amounts of bit errors. 
Figs.~6 and~7  show statistical measurements of our test chip in various conditions, and highlight the energy benefits of accepting more errors.
Finally, operating devices in high BER regimes allows using  conditions where they feature outstanding endurance. Fig.~8 for example shows endurance measurements of two devices programmed with low RESET voltages ($1.5V$). An endurance of more than $10^{10}$ cycles is seen, which is particularly high for such technology. This type of high cyclability opens the way to the possibility of training neural networks on chip, as seen in the results reported in \cite{hirtzlin2019hybrid}. 
A more detailed analysis of the energy benefits (which can reach a factor ten) of embracing bit errors in RRAM-based BNNs, and of the associated endurance benefits, is presented in \cite{hirtzlin2019digital}.

The strategy reported in this work is not limited to RRAM, and can be applied to other types of memories. Fig.~9 shows, based on neural network simulation, the energy that could be saved by varying the programming time of 28~nm Spin Torque Magnetoresistive RAM (ST-MRAM) using the same approach as the one presented here. We see that high energy savings can be achieved.  The methodology and model for obtaining these results are presented in \cite{hirtzlin2019implementing}.


\section{Conclusion}

Digital computing usually assumes and requires perfection in the memory bits, and this accuracy comes at important costs in terms of area and energy consumption. 
In contrast, neuromorphic circuits, including  fundamentally digital ones such as binarized neural networks can get away with imperfect memory cells. 
In this work, we use a differential approach to reduce errors and to be compatible with in or near-memory computing. This differential coding, in combination with the inherent tolerance of neural network, shows that it is possible on one side to embrace memories as ``non ideal'' without noticeable impact on neural network accuracy, and on the other side to get important benefits in terms of tuning of operating conditions (endurance, energy), opening the way to on-chip learning.

\bibliography{IEEEabrv,Mabibliotheque}
\bibliographystyle{IEEEtran}
\end{document}